\documentclass[reprint, amsmath,amssymb, preprintnumbers,aps,nofootinbib,superscriptaddress]{revtex4-1}
\pdfoutput=1
\usepackage{graphicx}
\usepackage{amssymb}
\usepackage{amsmath, amsthm, amssymb}
\usepackage{dcolumn}
\usepackage{bm}
\usepackage{slashed}
\usepackage{rotating}
\usepackage[hidelinks]{hyperref}
\usepackage{array}
\usepackage{color}
\usepackage{multirow}
\usepackage{xcolor}
\usepackage[T1]{fontenc}
\usepackage{tocloft}
\usepackage[normalem]{ulem}
\usepackage{mathrsfs}
\usepackage{units}
\usepackage{wasysym}
\usepackage[shortlabels]{enumitem}

\newcommand{\nn}{\nonumber}

\newcommand{\be}{\begin{equation}}
\newcommand{\ee}{\end{equation}}
\newcommand{\bea}{\begin{eqnarray}}
\newcommand{\eea}{\end{eqnarray}}

\newcommand{\p}{\partial}
\newcommand{\cO}{\mathcal{O}}

\newcommand{\ord}[1]{\cO(#1)}
\def\g{\gamma}
\renewcommand{\a}{\alpha}
\renewcommand{\b}{\beta}
\newcommand{\hc}{\text{h.c.}}
\DeclareMathOperator{\diag}{diag}

\begin{document}

\title{No-go limitations on UV completions of the Neutrino Option}

\author{Ilaria Brivio}
\affiliation{Institut f\"ur Theoretische Physik, Universit\"at Heidelberg, Philosophenweg 16, DE-69120 Heidelberg, Germany}
\author{Jim Talbert}
\affiliation{Niels Bohr Institute, University of Copenhagen, Blegdamsvej 17, DK-2100 Copenhagen, Denmark}
\author{Michael Trott}
\affiliation{Niels Bohr Institute, University of Copenhagen, Blegdamsvej 17, DK-2100 Copenhagen, Denmark}
\affiliation{CERN Theory Division, CH-1211 Geneva 23, Switzerland}

\begin{abstract}
We discuss the possible origin of the Majorana mass scale(s) required for the ``Neutrino Option''
where the electroweak scale is generated simultaneously with light neutrino masses in a type-I seesaw model,
by common dimension four interactions.
We establish no-go constraints on the perturbative generation of the Majorana masses required
due to global symmetries of the seesaw Lagrangian.

\end{abstract}
\maketitle

\section{Introduction and Motivation}
\label{sec:INTRO}
Amongst the outstanding theoretical issues of the Standard Model (SM), the origin(s) of neutrino masses and
the electroweak (EW) scale rank amongst the most pressing. Experiment has established
that at least two neutrinos are massive, and that the Higgs mass $m_h \simeq 125 \,{\rm GeV} \gg \delta m_\nu = m_{\nu_1} - m_{\nu_2}$.
These experimental facts, combined with the sensitivity of the Higgs
mass to high mass scale threshold corrections, are challenges to any ultraviolet (UV) completion of the SM
that seeks to explain the observed mass scales.
Although they are most often addressed independently, attempts at unified explanations of these observed masses are of great interest.

An interesting and minimal possibility is that both the mass scales, $m_h$ and $m_{\nu_1}\sim m_{\nu_2}$,
are generated simultaneously in a minimal extension of the SM
from an underlying Majorana scale. Ref.~\cite{Brivio:2017dfq} showed that this scenario, dubbed the ``Neutrino Option'',
can be realized within the simplest type-I seesaw model~\cite{Minkowski:1977sc,GellMann:1980vs,Mohapatra:1979ia,Yanagida:1980xy,Schechter:1980gr}. 
This approach has been shown to be compatible with the observed neutrino mass and mixing patterns~\cite{Brivio:2018rzm} and resonant leptogenesis~\cite{Brivio:2019hrj,Brdar:2019iem}.
It admits UV completions where approximate scale invariance plays an important role~\cite{Brdar:2018vjq,Brdar:2018num,Brivio:2019hrj} and also non-perturbative ones, e.g. with strongly-interacting hidden sectors that add viable Dark Matter candidates to the spectrum~\cite{Aoki:2020mlo}, and in certain string compactifications~\cite{Talbert:2020mny}.
In this setup, the traditional Higgs mass hierarchy problem translates into a quest for a UV origin of the underlying
Majorana mass scale, with the required pattern of threshold corrections.

In this paper, we study possible UV completions of the Neutrino Option.
We use the minimal scenario that the Majorana scale required by the Neutrino Option
is generated perturbatively from a deep-UV scale associated with a very heavy Majorana state.
We show how symmetries of the seesaw Lagrangian, and the specific parameter space required in the Neutrino Option,
makes minimal model scenarios relying on one-loop corrections run up against seesaw model symmetry constraints.
We discuss minimal extensions that might evade our conclusions.
The primary results we present are some no-go constraints for UV-completing the Neutrino Option
in the minimal setups we consider.

\section{The Neutrino Option}
\label{sec:NeutrinoOption}

Consider the Type-I seesaw model, where the SM is extended with three right-handed (RH) spinors $N_{R,p}$, with $p = \lbrace 1, 2, 3 \rbrace$. The field $N_p$ defined by~\cite{Bilenky:1980cx,Broncano:2002rw}
\begin{equation}
N_{p} = e^{i \theta_p /2} N_{R,p} + e^{-i \theta_p /2} (N_{R,p})^c\,,
\end{equation}
with $\theta_p$ an arbitrary phase, satisfies the Majorana condition $N_p = N_p^c$~\cite{Majorana:1937vz}. Here the superscript $c$ denotes charge conjugation: $\psi^c = C \bar \psi^T$, with $C= - i \gamma_2 \gamma_0$ in the chiral basis for the $\gamma_i$.
The seesaw Lagrangian is \footnote{Chiral projection and charge conjugation do not commute. In this paper $\psi_{L/R}^c$ denotes a field chirally projected and subsequently charge conjugated.}
\begin{align}
\label{eq:LagSS}
\mathcal{L}_N &=\, \frac{1}{2}\left( \bar N_p i \slashed{\partial} N_p
-\bar N_p M_{pr} N_r\right)
- \left[\bar N_p\,  \omega_{p\b}\, \tilde H^\dag l_\b+\hc \right]
\nn\\
&=\,\frac{1}{2}\left[\bar N_{R,p} i\slashed{\p}N_{R,p}+\overline{N^c_{R,p}}\, i\slashed{\p}N_{R,p}^c\right]
\\
&- \left[\frac{1}{2}e^{-i\theta_p}\,\bar N_{R,p} \,M_{pr} N_{R,r}^c +
e^{-i\theta_p/2}\,
\bar N_{R,p}\,  \omega_{p\b}\, \tilde H^\dag l_\b
 + \hc\right] \,,
\nn
\end{align}
where $l$ is the left-handed (LH) SM lepton doublet and $\b = \lbrace 1,2,3 \rbrace$ its associated flavor index. The resulting mass matrix is symmetric:  $(e^{-i\theta}M)= (M^Te^{-i\theta})$ .

\begin{figure}
\includegraphics[width=90mm]{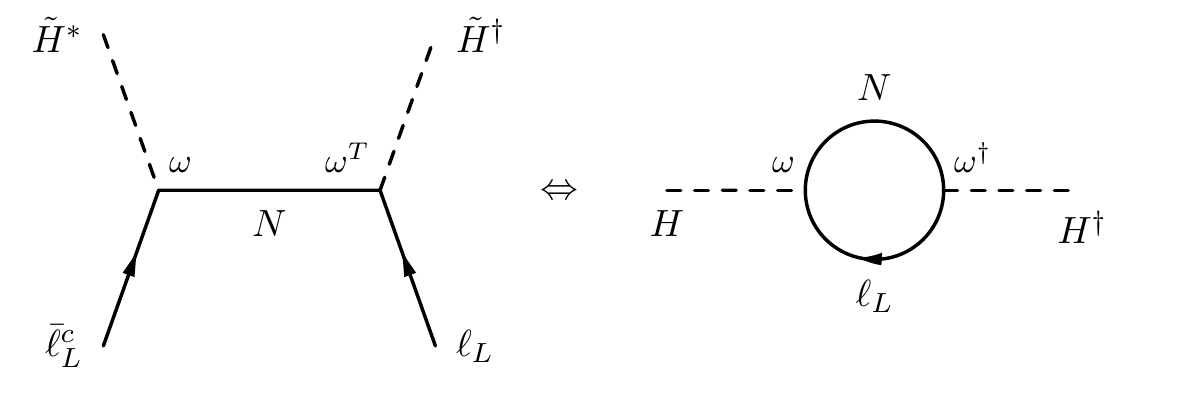}
\caption{One-loop threshold corrections generating the EW scale in the Neutrino Option. The one loop diagram
is linked to neutrino mass generation by connecting the lepton line.}
\label{fig:match}
\end{figure}

The phenomenology of $\mathcal{L}_{SM} +\mathcal{L}_N$ at $p^2 \ll M_p^2$ has the $N_p$ fields integrated out in sequence
and matched to the SMEFT. The tree-level matching is known up to dimension seven
\cite{Weinberg:1979sa,Wilczek:1979hc,Broncano:2002rw,Broncano:2003fq,Broncano:2004tz,Abada:2007ux,Elgaard-Clausen:2017xkq}
and at dimension five one finds
\begin{align}
\mathcal{L}^{(5)} &= \frac{c_{\a\b}^{(5)}}{2}
\left(l_\a^T \tilde H^*\right)C\left(\tilde H^\dag l_\b\right)+\hc
\,,
\\
c_{\a\b}^{(5)} &= \left(\omega^T M^{-1}\omega\right)_{\a\b}\,,
\end{align}
One-loop matching introduces sub-leading corrections to $c^{(5)}$ and {\it necessarily induces threshold matching
contributions (c.f. the diagrams in Fig. \ref{fig:match}) to the SM Higgs mass from the same interactions}~\cite{Brivio:2017dfq,Brivio:2018rzm,Brivio:2019hrj}:
\begin{equation}
V(H) = -\frac{m_{h0}^2 + \Delta m_h^2}{2}\, H^\dagger H + (\lambda_0+\Delta\lambda)\, (H^\dagger H)^2\,.
\end{equation}
Here $m_{h0},\lambda_0$ are the ``bare'' parameters defining the classical scalar potential
at $\mu\simeq M$ and $\Delta m_h^2,\,\Delta\lambda$ are the loop-induced threshold corrections.
Assuming a nearly conformal classical Lagrangian implies $m_{h0}\simeq0$, while $\lambda_0$ is free and generally
of perturbative size. As $\Delta\lambda\propto\omega^4$, this contribution is typically negligible for perturbative Yukawa couplings $|\omega_{p\b}|< 1$. On the other hand\footnote{This expression is derived in the basis where $M$ is diagonal.}
\begin{equation}
\label{eq:Higgsthresh}
\Delta m_h^2 = \frac{1}{8\pi^2}{\rm Tr}\left(\omega^\dagger M^2 \omega\right)\,,
\end{equation}
is generally large and directly sensitive to the Majorana mass scale. This contribution has been long known and is a direct representation of the hierarchy problem in the seesaw model, see e.g. Refs.~\cite{Vissani:1997ys,Casas:1999cd,Bambhaniya:2016rbb}.
The key idea of the Neutrino Option is that, taking $m_{h0}\simeq 0$, $\Delta m_h^2$ can be interpreted as a radiatively generated Higgs mass.
It is interesting that Fermi statistics in this minimal setup fixes the sign of this threshold correction to be negative,
as required
to induce a low scale Higgs'd phase of electroweak symmetry breaking in the SM with massive neutrinos.
Requiring that both the observed EW scale and neutrino mass-squared differences are accommodated identifies the parameter space~\cite{Brivio:2017dfq,Brivio:2018rzm}\footnote{The requirement $|\omega|\sim\unit{TeV}/M$ stems from $\Delta m_h^2\simeq(\unit[100]{GeV})^2$. Inserting it in the expression for light neutrino masses and requiring $m_\nu\gtrsim\unit[0.01]{eV}\sim\Delta m_{\astrosun}$ identifies the upper limit on $M$. }
\begin{align}
\label{eq:parameterspace}
M &\lesssim \unit[10^4]{TeV} \simeq  \unit[10]{PeV}\,.
&
|\omega| &\simeq \frac{\unit{TeV}}{M}\,.
\end{align}
Requiring successful resonant leptogenesis introduces an additional lower limit on the Majorana scale~\cite{Brivio:2019hrj}
\begin{equation}\label{eq:parameterLG}
M\gtrsim \unit[1]{PeV}\,.
\end{equation}
Finer structure of the allowed parameter space can be identified specifying the neutrino mixing parameters and CP violating phases.
These exact results are sensitive to the top quark mass, the order of the RG equations used,
and the details of the seesaw model, such as the number of RH neutrinos introduced and the structure of the $M$ matrix.
On the other hand, the orders of magnitude in~Eqs.~\eqref{eq:parameterspace},~\eqref{eq:parameterLG}
have a negligible dependence on these choices.

\subsection{Symmetries of the seesaw Lagrangian}
The  Lagrangian has the following global symmetries:
\begin{itemize}
\item The kinetic term of the $N$ fields respects a global $\rm U(3)_N$ flavour symmetry, that can be decomposed into
$\rm U(1)_{N,3}\times SU(3)_N$, where the $\rm U(1)_{N,k}$ term represents an $N$-lepton number under which $k$ flavors transform.
The kinetic term of the $l$ doublet has a $\rm U(3)_l=U(1)_{l,3}\times SU(3)_l$ symmetry.

\item Discrete and continuous symmetries are associated to massive and massless $N_p$ states.
With $n$ non-zero eigenvalues, the Majorana mass term breaks the $\rm U(3)_N$ down
to
\begin{align}
&U(1)_{N,2}\times SU(2)_N\times \mathbb{Z}_2
&
(n&=1)\,,
\\
&U(1)_N\times  \mathbb{Z}_2 \times \mathbb{Z}_2
&
(n&=2)\,,
\\
&\mathbb{Z}_2 \times \mathbb{Z}_2
&
(n&=3)\,.
\end{align}
The Klein four group $\mathbb{Z}_2 \times \mathbb{Z}_2$ is the maximal discrete symmetry of $M_{pr}$
\cite{Lam:2007qc,King:2013eh}.
\item
The neutrino Yukawa term preserves only the diagonal lepton number $\rm U(1)_{l+N}\supset U(1)_{l,3}\times U(1)_{N,3}$ and breaks explicitly all the remaining flavor symmetries.

\end{itemize}

The $\mathbb{Z}_2$ symmetries, if preserved, protect the Higgs mass against corrections proportional to the associated Majorana mass.
Consider for instance a case where $M$ is diagonal with only $M_{33} \neq 0$.  The associated preserved mass-eigenstate $\mathbb{Z}_2$ transformation can be represented in flavour space by
\begin{equation}
\label{eq:Tmass}
N_p \longrightarrow T_{pr}\,N_r \,, \quad\text{with}\quad T_{pr} = \diag\left( 1,1,-1 \right)\,.
\end{equation}
Invariance of the Yukawa terms then implies
\begin{align}
\nonumber
&\bar N \,\omega \, \tilde{H}^\dagger \, l
\;\overset{!}{=}\;
\bar N\,T^\dagger \,\omega \, \tilde{H}^\dagger \, l \
\\
\label{eq:Z2kill}
&\Rightarrow \; T^{\dagger}\,\omega = \omega
\;\Rightarrow\;
\omega_{3\b}\equiv 0\,.
\end{align}

Comparing to Eq.~\eqref{eq:Higgsthresh}, this indicates that an exact $\mathbb{Z}_2$ symmetry forbids contributions to $\Delta m_h^2$ from  $M_{33}$.  This also occurs if there are two heavy mass states, both of which respect an associated $\mathbb{Z}_2$.

\section{Perturbative generation of the Majorana mass}
\label{sec:Perturbative}

The origin of the scale $M\sim 10 \, {\rm PeV}$ is the main theoretical question left open in the formulation of the Neutrino Option.
Even though the model only contains interactions up to dimension four, a generation mechanism for $M$ is required to ensure the validity
of the key assumption in this construction, namely that the Majorana mass term is generated
without other large threshold corrections, and also in a manner that dominantly breaks
the approximate classical scale invariance in the rest of the Lagrangian.

On very general grounds, a successful generation mechanism should have the following properties:
\begin{enumerate}[(i)]
\item It is required to generate at least 2 $M$ eigenvalues at the PeV scale. $n\geq 2$ is required  for consistency with the 2 non-zero mass splittings observed in the light neutrino spectrum.
\label{assum:pev}
\item The Higgs mass term does not receive additional large threshold contributions besides those in Eq.~\eqref{eq:Higgsthresh}.
This condition can be associated with approximate classical scale invariance.
\label{assum:mh}
\item From an EFT perspective, any UV completion of the seesaw Lagrangian generally extends it with higher dimensional operators. Although most of these can be safely neglected in the phenomenology of the Neutrino Option, certain structures, such as $(\bar N N)(H^\dag H)$,  can potentially destabilize
the Higgs mass, and will not necessarily be protected by discrete symmetries.
The absence of these operators was an implicit assumption in the original formulation of the Neutrino Option,
and they should not be generated with unsuppressed Wilson coefficients.

\label{assum:uveft}
\item The RGE running of the Higgs and neutrino parameters is not spoiled by new light BSM states.
\label{assum:rge}
\item The parameter space does not rely on strong tunings. This latter condition can be associated with technical naturalness,
or be purely aesthetic. Avoiding parameter tuning directly leads to the idea that heavy UV mass scales should be associated with
Fermionic states avoiding massive Bosonic states, that can couple to $H^\dagger H$.
\label{assum:notuning}
\end{enumerate}

Here we consider the possibility that the PeV scale originates perturbatively through threshold corrections or RG
evolution from some deeper UV Majorana scale, which is arguably the minimal scenario,
and a very simple possibility, because then such perturbations arise due to loop effects in the seesaw model itself.
Potential one-loop corrections, and low scale mass terms scale as
\begin{equation}
\delta_M^{(1)} =\frac{|\omega|^2}{16\pi^2}\, M_{UV}\,,
\quad\rightarrow\quad
\epsilon = \frac{|\omega|^2}{16\pi^2}\,,
\end{equation}
which, interestingly, is in the desired ballpark for values of the Yukawa coupling that lie within the
phenomenologically allowed range for the Neutrino Option for some interesting $\rm UV$ scales
\begin{equation}
\label{eq:omegarange}
\delta_M^{(1)} \simeq \unit{PeV} \quad\text{ for }\quad
\begin{cases}
|\omega|\simeq 10^{-4}\,, & (M_{UV} \simeq M_{GUT})
\\
|\omega|\simeq 10^{-5.5}\,. & (M_{UV}\simeq  M_{Pl})
\end{cases}
\end{equation}
Following this numerical coincidence, a minimal hypothesis is that the a UV mechanism
that is flavor-blind leads to the democratic texture
\begin{equation}
\label{eq:degenM}
M = \frac{M_{UV}}{3}
\begin{pmatrix}
1 & 1 & 1 \\
1 & 1 & 1 \\
1 & 1 & 1
\end{pmatrix}\,,
\end{equation}
that, once diagonalized, leaves two massless eigenstates, and one massive state
\begin{equation}\label{eq.M00}
M =
\begin{pmatrix}
0& & \\ & 0& \\ & & M_{UV}
\end{pmatrix}\,.
\end{equation}
In such a UV scenario, a super-heavy Majorana mass scale $M_{UV}$ is assumed to emerge from high scale dynamics.
The democratic flavour blind mass generation mechanism is the minimal possibility as
the Majorana fields carry no (SM) quantum numbers. It has been argued 
that such a mass matrix
is a straightforward expectation when the mass generation is associated with gravity ~\cite{Akhmedov:1992hh}.

In the presence of perturbations  of order $\epsilon\ll 1$ to the texture in Eq.~\eqref{eq:degenM},
the zero eigenvalues are generally lifted and replaced by $\ord{\epsilon\, M_{UV}}$ quantities.
 For $\epsilon\sim 10^{-13} (10^{-10})$ and $M_{UV}= M_{Pl} (M_{GUT})$, this would successfully identify the PeV scale.

Unfortunately, this scenario is not realized at the one-loop level in the most minimal setup we consider.
In order for a Majorana mass eigenvalue to be non-zero, its associated lepton
number must be violated by two units.
Given $\mathcal{L}_N$ with the mass matrix in Eq.~\eqref{eq.M00},
no tree or one-loop diagram topology exists with this property,
see Refs.~\cite{Casas:1999tp,Casas:1999tg,Brivio:2018rzm}. The same is true for mass matrices with $n=2$.
This implies that the texture-zero(es) are preserved by both threshold corrections and RGE running at one-loop
in type-I seesaw models.

Assuming a heavy scale $M_{UV}$, and pursuing this minimal scenario further, one is then left with two perturbative alternatives:
\begin{enumerate}[A.]
\item $n\geq2$ eigenvalues of order $M_{UV}$ are present, and the one-loop RGE running induces a large suppression that reduces them to the PeV scale.

\item starting from $n=1$ non-zero eigenvalue, the PeV scale is generated radiatively  at 2 or more loops.
\end{enumerate}

Due to the simultaneous requirement of $L$-violating and $\mathbb{Z}_2$-preserving interactions
(from neutrino and Higgs mass considerations, respectively), neither of these two possibilities turns out to be
consistent with the minimal extension of the Neutrino Option scenario we consider, as we discuss in the next subsections.

\subsection{One-loop RGE flow}
\label{sec:oneloop}

Consider the one-loop case, where the RG equation for the Majorana mass term is
\cite{Casas:1999tp,Casas:1999tg,Antusch:2005gp,Antusch:2003kp}
\begin{equation}
\label{eq:MajRun}
16 \pi^2 \mu \frac{d M}{d\mu} = \left(\omega \omega^\dagger\right) M +  M \left(\omega \omega^\dagger\right)^T\
 \equiv \mathcal{R}\,,
\end{equation}
and $\mathcal{R}=\mathcal{R}^T$.
Diagonalizing, the mass eigenvalues evolve multiplicatively~\cite{Casas:1999tp,Casas:1999tg}:
\bea
M_p(\mu) &=& \gamma_p(\mu,\mu_0) M_p(\mu_0)\,,\\
\gamma_p (\mu, \mu_0) &\sim& 1 +\frac{\omega^2}{16\pi^2} \ln\left[\frac{\mu}{\mu_0}\right]\,.
\eea

When all the $\mathbb{Z}_2$ symmetries associated to the massive $N$ states are preserved, $\mathcal{R} \equiv 0$,
and equivalently $\gamma_p\equiv 1$.
This can easily be seen by using Eq.\eqref{eq.M00} and Eq.\eqref{eq:Z2kill} in Eq.\eqref{eq:MajRun}.
Hence, for the running to occur, the $\mathbb{Z}_2$ needs to be at least softly broken.
In this case, for the PeV scale to emerge from RG running one requires
\begin{equation}
\label{eq:epsilontune}
\gamma_p(\unit{PeV}, M_{UV})= \epsilon\simeq \frac{\unit{PeV}}{M_{UV}}\sim 10^{-10}-10^{-13}\,,
\end{equation}
which immediately implies that the radiative contribution $\omega^2/16\pi^2 \ln \left[ \mu/\mu_0 \right]$ must be tuned.

\subsection{Higher perturbative orders}

\begin{figure}
\includegraphics[width=6cm]{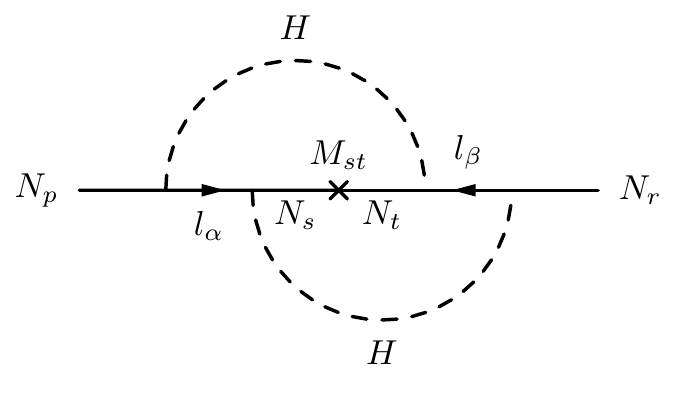}
\caption{$\Delta L=2$ two-loop diagram that allows a perturbative generation of a Majorana mass term $M_{pr}$.}
\label{fig:twoloop}
\end{figure}

The generation of $\Delta L=2$ amplitudes is possible at two-loops in the type-I seesaw,
via diagrams such as the one in Fig~\ref{fig:twoloop}, see
Refs.~\cite{Leung:1983ti,Petcov:1984nz,Ibarra:2018dib,Ibarra:2020eia}. These diagrams generally contribute to all entries
of the Majorana mass matrix $M$, including the off-diagonals.

Two-loop radiative corrections in the seesaw scale as~\cite{Ibarra:2018dib}
\begin{equation}
\delta_{M,pr}^{(2)} \sim \frac{(\omega\omega^\dagger)_{p3}(\omega\omega^\dagger)_{r3}}{256\pi^4}\, M_{UV}\,.
\end{equation}
As in the one-loop RGE case, $\delta_M^{(2)}\neq 0$ only
if $\omega_{3\beta}\neq 0$, i.e when the $\mathbb{Z}_2$ symmetry associated to the massive state ($N_3$) is broken,
which leaves the Higgs mass term unprotected.
Indeed,  it can be checked that both threshold corrections and RG equations at two-loops only contain the flavor structures~\cite{Ibarra:2018dib,Ibarra:2020eia,Antusch:2002ek}:
\begin{align}
\omega \omega^\dagger M\,, \quad
\omega \omega^\dagger M (\omega \omega^\dagger)^T\,, \quad
(\omega \omega^\dagger)(\omega \omega^\dagger )M\,,
\end{align}
and their transposes,
that vanish identically in the $\mathbb{Z}_2$ symmetric limit. This statement is independent of $n$.

This leads to a tension between the requirements~\ref{assum:pev}~and~\ref{assum:mh} above:
in order to have $\delta_M^{(2)}\sim\unit{PeV}$ with  $M_{UV}=M_{Pl}(M_{GUT})$, the Yukawa couplings should be $(\omega\omega^\dag)_{p3} \lesssim 1(5)\times 10^{-4}$.
Inserting this value in Eq.~\eqref{eq:Higgsthresh}, the contribution to $\sqrt{\Delta m_h^2}$ from  $M_{33}=M_{UV}$ is $M_{UV}\sqrt{(\omega\omega^\dag)_{33}/8\pi^2} \simeq \unit[10^{16}]{GeV}$.

On the other hand, assuming that the $\mathbb{Z}_2$ is only very softy broken in order to protect $\Delta m_h^2$,
leads to $\delta_M^{(2)}$ of sub-eV size.  While such masses may be interesting for low-energy
phenomenology (e.g. for sterile neutrino dark matter studies), they do not account for the preferred
coupling ranges of the Neutrino Option. These symmetry and scaling arguments
hold both for two-loop threshold and RG contributions, and this tension persists at higher
perturbative orders.

\vskip 1em

\section{Comments on extended model field content}
The conclusions in previous sections
are drawn under the assumption of no additional BSM field content beyond
that of $\mathcal{L}_{SM}+ \mathcal{L}_N$.
A perturbative generation mechanism from the deep UV involving more fields
must still avoid the generic global symmetry arguments;
a mass generation mechanism must provide radiative generation of a $\Delta L = 2$ Feynman
diagram to lift zero Majorana eigenvalues, or radiatively generate lower scales from existing $M_{UV}$ eigenvalues.
One must simultaneously preserve a symmetry protection to control threshold corrections from
heavy Fermionic or Bosonic states, to prevent large Higgs threshold corrections.

\begin{figure}
\includegraphics[width=70mm]{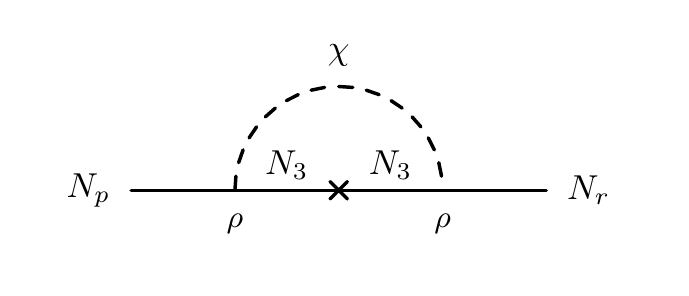}
\caption{A sample one-loop diagram that can radiatively generate a new non-zero Majorana neutrino mass, starting from a unique non-vanishing eigenvalue $M_{33}$. The state $\chi$ is an unspecified boson whose properties can be partially derived, see the main text.}
\label{fig:oneloopsuccess}
\end{figure}

Consider adding a generic Boson field to the minimal setup in $\mathcal{L}_{SM}+ \mathcal{L}_N$.
Since $N$ is a majorana field, the allowed $\bar NN\chi$ couplings up to dimension four are
$$\frac12\bar N\left[\rho_S\, \chi_S + i \rho_P\, \g_5 \chi_P\right]N\,.$$
Here $\chi_S,\,\chi_p$ are real scalar and pseudoscalar couplings respectively.
For gauge invariance of the $\bar NN \chi$ coupling, $\chi$ must be a $\rm SU(3)_C\times SU(2)_L\times U(1)_Y$ singlet.
In general such couplings are off diagonal in $N_p$ flavour space.
In the limit of vanishing external momenta, the graph in Fig.~\ref{fig:oneloopsuccess} scales as
\begin{equation}
\frac{(\rho_S^2+ \rho_P^2) \, M_{33}}{64\pi^2}\left[1+ \frac{1}{M_{33}^2-m_\chi^2}
\left(M_{33}^2 \log\frac{\mu^2}{M_{33}^2} - m^2_\chi \log\frac{\mu^2}{m_\chi^2}\right)\right]
\end{equation}
inducing lower scale mass eigenvalues in the $M_{pr}$ matrix from the UV scale mass $M_{33}$.
In addition, an anapole/Zeldovich \cite{anapole} coupling is allowed for a vector field coupling to a Majorana
bilinear
$$\frac{1}{2 \Lambda^2}\bar N\left[\rho_V \, \g_\mu \g_5 \right]N \partial_\nu \chi_V^{\nu \mu}\,.$$
In the case of the SM, $\chi_V^{\nu \mu} = B^{\mu \nu}$ is generated at one
loop if the SM states are massive with a closed Higgs and charged lepton in the loop.
In this case, the couplings $\omega$ associated with the numerical coincidence in Eq.~\eqref{eq:omegarange} are present.
However, in the minimal extension of the Neutrino Option we consider,
these states are effectively massless, and the one loop diagram vanishes in dimensional regularization as the integrals are scaleless.
The anapole moment can be induced by UV field content leading to induced masses
\begin{equation}
\frac{3 \, \rho_V^2 \, M_{33}^5}{64\pi^2 \,\Lambda^4}\left[
1 + r + r^2+
\frac{
\log\frac{\mu^2}{M_{33}^2}
-
r^3\log\frac{\mu^2}{m_\chi^2}
}{1-r}\right],
\end{equation}
with $r=m_\chi^2/M_{33}^2$.
In general, there is no association of $\rho_V^2$ with the numerical coincidence in Eq.~\eqref{eq:omegarange}
and the combination  $\rho_V^2  M_{33}^4/\Lambda^4$ can be chosen to induce a mass hierarchy.
Further, in the case of a gauged $\rm U(1)$ field  $\chi_V^{\nu \mu}$ the Higgs portal coupling
is through $\lambda' \chi_V^{\nu \mu} \chi_V^{\nu \mu} H^\dagger H/\Lambda^2$.
The usual hierarchy problem is present in the case of the BSM induced anapole moment, and in the case of a scalar field $\chi_{S/P}$.
$m_{\chi}$ itself is not protected against contributions of $\mathcal{O}(M_{33})$ as a result of the $\mathbb{Z}_2$ symmetry constraints.

Seeking to associate the Bosonic coupling with $\omega$ and also cancel the threshold contribution to the
Higgs mass leads to a minimally-supersymmetric SM (MSSM) extended with singlets $N$ in the UV.
The Higgs mass threshold correction is then canceled, but the one-loop analogue to Eq.~\eqref{eq:MajRun}
is given in Ref.~\cite{Antusch:2002ek}. No new $\Delta L = 2$ contribution is added
and the seesaw RGE equation is merely rescaled by an overall factor compared to the SM.
The conclusions of the previous section hold.

One can also consider models where $L$ violation resides in BSM couplings, rather than relying on the initial $M_{33}$ term. However, achieving the desired symmetry-breaking and conserving patterns is not any easier in these scenarios.
Consider for example leptoquarks, which can have either ($B+L$) conserving and ($B-L$) violating couplings or vice-versa. Because they necessarily carry other quantum numbers (most importantly color charge), the leptoquark lines always need to be closed with conjugate vertices. This means that, in a one-loop diagram, $B$ and $L$ violating terms always compensate each other. In fact, independently of their quantum numbers, leptoquarks can only generate a Majorana mass at two loops~\cite{Babu:2010vp,Dorsner:2016wpm}.
Any two-loop diagram is highly suppressed and not directly associated with the numerical coincidence in Eq.~\eqref{eq:omegarange}
to achieve $M_p \sim$ PeV.
\\
\\
\section{Summary and Outlook}
\label{sec:CONCLUDE}
We have examined minimal extensions of the Neutrino Option setup to induce $M_p \sim$ PeV Majorana masses
from a deep UV mass scale. Such an approach is primarily motivated due to the numerical coincidence identified in
Eq.~\eqref{eq:omegarange}.
Such perturbative mass generation mechanisms must overcome the combined symmetry constraints
of required $L$ violation while suppressing threshold contributions to the Higgs mass. A natural symmetry protection mechanism
relies on the $\mathbb{Z}_2$ symmetries present in the $\mathcal{L}_{SM}+ \mathcal{L}_{N}$ Lagrangian itself.
Such $\mathbb{Z}_2$  based symmetry protection, in the minimal setups we considered,
blocks the generation $M_p \sim$ PeV Majorana masses from the deep UV scales and forbids a natural realization of the
numerical coincidence identified in Eq.~\eqref{eq:omegarange}. It is possible that more extended model building could overcome
this challenge.

\section*{Acknowledgements}
J.T. and M.T. acknowledge support from the Villum Fund, project number 00010102.
J.T. thanks Ivo de Medeiros Varzielas for helpful comments regarding explicit flavour-symmetry breaking. I.B. thanks Peter Denton for useful discussions.

\bibliographystyle{apsrev4-1}
\bibliography{biblioPert}
\end{document}